\documentclass[preprint,showpacs,preprintnumbers,amsmath,amssymb,superscriptaddress]{revtex4}

\usepackage{CJK}

\usepackage{graphicx}
\usepackage{dcolumn}
\usepackage{bm}
\usepackage{tabularx}

\begin{document}
\begin{CJK*}{GBK}{song}

\title{Spectroscopic factors for low-lying $^{16}$N levels and the astrophysical $^{15}$N($n$,\,$\gamma$)$^{16}$N reaction rate}

\author{B. Guo}\thanks{guobing@ciae.ac.cn}\affiliation{China Institute of Atomic Energy, P.O. Box 275(10),
Beijing 102413, China}
\author{Z. H. Li}\affiliation{China Institute of Atomic Energy, P.O. Box 275(10),
Beijing 102413, China}
\author{Y. J. Li}\affiliation{China Institute of Atomic Energy, P.O. Box 275(10),
Beijing 102413, China}
\author{J. Su}\affiliation{China Institute of Atomic Energy, P.O. Box 275(10),
Beijing 102413, China}
\author{D. Y.
Pang}\affiliation{School of Physics and Nuclear Energy
Engineering, Beihang University, Beijing 100191,
China}\affiliation{International Research Center for Nuclei and
Particles in the Cosmos, Beihang University, Beijing 100191,
China}
\author{S. Q. Yan}\affiliation{China Institute of Atomic Energy, P.O. Box 275(10),
Beijing 102413, China}
\author{Z. D. Wu}\affiliation{China Institute of Atomic Energy, P.O. Box 275(10),
Beijing 102413, China}
\author{E. T.
Li}\affiliation{College of Physics Science and Technology,
Shenzhen University, Shenzhen 518060, China}
\author{X. X. Bai}\affiliation{China Institute of Atomic Energy, P.O. Box 275(10),
Beijing 102413, China}
\author{X. C. Du}\affiliation{China Institute of Atomic Energy, P.O. Box 275(10),
Beijing 102413, China}
\author{Q. W. Fan}\affiliation{China Institute of Atomic Energy, P.O. Box 275(10),
Beijing 102413, China}
\author{L. Gan}\affiliation{China Institute of Atomic Energy, P.O. Box 275(10),
Beijing 102413, China}
\author{J. J.
He}\affiliation{Institute of Modern Physics, Chinese Academy of
Sciences (CAS), Lanzhou 730000, China}
\author{S. J. Jin}\affiliation{China Institute of Atomic Energy, P.O. Box 275(10),
Beijing 102413, China}
\author{L. Jing}\affiliation{China Institute of Atomic Energy, P.O. Box 275(10),
Beijing 102413, China}
\author{L. Li}\affiliation{Institute of Modern Physics, Chinese Academy of
Sciences (CAS), Lanzhou 730000, China}
\author{Z. C. Li}\affiliation{China Institute of Atomic Energy, P.O. Box 275(10),
Beijing 102413, China}
\author{G. Lian}\affiliation{China Institute of Atomic Energy, P.O. Box 275(10),
Beijing 102413, China}
\author{J. C. Liu}\affiliation{China Institute of Atomic Energy, P.O. Box 275(10),
Beijing 102413, China}
\author{Y. P. Shen}\affiliation{China Institute of Atomic Energy, P.O. Box 275(10),
Beijing 102413, China}
\author{Y. B. Wang}\affiliation{China Institute of Atomic Energy, P.O. Box 275(10),
Beijing 102413, China}
\author{X. Q. Yu}\affiliation{Institute of Modern Physics, Chinese Academy of
Sciences (CAS), Lanzhou 730000, China}
\author{S. Zeng}\affiliation{China Institute of Atomic Energy, P.O. Box 275(10),
Beijing 102413, China}
\author{L. Y. Zhang}\affiliation{Institute of Modern Physics, Chinese Academy of
Sciences (CAS), Lanzhou 730000, China}
\author{W. J. Zhang}\affiliation{China Institute of Atomic Energy, P.O. Box 275(10),
Beijing 102413, China}
\author{W. P. Liu}\affiliation{China Institute of Atomic Energy, P.O. Box 275(10),
Beijing 102413, China}

\date{\today}

\begin{abstract}

\noindent\textbf{Background:} Fluorine is a key element for
nucleosynthetic studies since it is extremely sensitive to the
physical conditions within stars. The astrophysical site to
produce fluorine is suggested to be asymptotic giant branch (AGB)
stars. In these stars the $^{15}$N($n$,\,$\gamma$)$^{16}$N
reaction could affect the abundance of fluorine by competing with
$^{15}$N($\alpha$,\,$\gamma$)$^{19}$F.

\noindent\textbf{Purpose:} The $^{15}$N($n$,\,$\gamma$)$^{16}$N
reaction rate depends directly on the neutron spectroscopic
factors of the low-lying states in $^{16}$N. Shell model
calculations and two previous measurements of the ($d,\,p$)
reaction yielded the spectroscopic factors with a discrepancy by a
factor of $\sim$2. The present work aims to explore these neutron
spectroscopic factors through an independent transfer reaction and
to determine the stellar rate of the
$^{15}$N($n$,\,$\gamma$)$^{16}$N reaction.

\noindent\textbf{Methods:} The angular distributions of the
$^{15}$N($^{7}$Li,\,$^{6}$Li)$^{16}$N reaction populating the
ground state and the first three excited states in $^{16}$N are
measured using a Q3D magnetic spectrograph and are used to derive
the spectroscopic factors of these states based on distorted wave
Born approximation (DWBA) analysis.

\noindent\textbf{Results:} The spectroscopic factors of these four
states are extracted to be 0.96\,$\pm$\,0.09, 0.69\,$\pm$\,0.09,
0.84\,$\pm$\,0.08 and 0.65\,$\pm$\,0.08, respectively. Based on
the new spectroscopic factors we derive the
$^{15}$N($n$,\,$\gamma$)$^{16}$N reaction rate.

\noindent\textbf{Conclusions:} The accuracy and precision of the
spectroscopic factors are enhanced due to the first application of
high-precision magnetic spectrograph for resolving the
closely-spaced $^{16}$N levels which can not be achieved in most
recent measurement. The present result demonstrates that two
levels corresponding to neutron transfers to the 2$s_{1/2}$ orbit
in $^{16}$N are not so good single-particle levels although
$^{15}$N is a closed neutron-shell nucleus. This finding is
contrary to the shell model expectation. The present work also
provides an independent examination to shed some light on the
existing discrepancies in the spectroscopic factors and the
$^{15}$N($n$,\,$\gamma$)$^{16}$N rate.

\end{abstract}

\maketitle


Fluorine ($^{19}$F, the only stable F isotope) is a crucial
element for nucleosynthetic studies because it is extremely
sensitive to the physical conditions within stars \cite{luc11}.
For a long time, the solar system was the only location of the
Galaxy with known fluorine abundance \cite{mey00}. The
astrophysical site to produce fluorine has always been a puzzle.
The possible scenario is suggested to be asymptotic giant branch
(AGB) stars \cite{for92,cri09}, core-collapse of Type II
supernovae \cite{woo90,woo02} and Wolf-Rayet stars \cite{mey00}.
Although the contribution of each source to the F evolution in the
Galaxy is still uncertain, the inclusion of all the three sources
is necessary to explain the observed Galactic evolution of
fluorine \cite{ren04}. In 1992, spectroscopic observations of
giant stars showed enhancements of $^{19}$F by factors of up to 30
with respect to solar abundances, providing the first evidence for
$^{19}$F nucleosynthesis in this site \cite{jor92}. This result
has been later supported by the large F enhancements found in
post-AGB stars \cite{wer05} and planetary nebulae
\cite{zha05,ots08} which are the progeny of AGB stars. However,
AGB model calculations have not yet been able to quantitatively
reproduce the highest values of observed $^{19}$F enhancements
\cite{lug04,cri09}. Reduction of nuclear reaction rate
uncertainties and a better understanding of the nucleosynthesis in
the partial mixing zone could help to clarify the discrepancies
\cite{lug04}. A detailed understanding of the nuclear reactions
occurring in AGB stars is desirable to simulate fluorine
production.

$^{19}$F is produced in the He intershell and then dredged up to
the surface of AGB stars. The production paths are
$^{14}$N($\alpha,\,\gamma$)$^{18}$F($\beta^+$)$^{18}$O($p,\,\alpha$)$^{15}$N($\alpha,\,\gamma$)$^{19}$F
and
$^{14}$N($n,\,p$)$^{14}$C($\alpha,\,\gamma$)$^{18}$O($p,\,\alpha$)$^{15}$N($\alpha,\,\gamma$)$^{19}$F.
Neutrons and protons are liberated by the
$^{13}$C($\alpha,\,n$)$^{16}$O and $^{14}$N($n,\,p$)$^{14}$C
reactions, respectively. Furthermore, the neutron- and
proton-induced reactions must also be taken into account although
$^{19}$F is produced in He-rich environments \cite{lug04,her03}.
The $^{15}$N($n$,\,$\gamma$)$^{16}$N reaction may play an
important role in determining the $^{19}$F abundance since it
competes with the $^{15}$N($\alpha,\,\gamma$)$^{19}$F reaction and
removes both neutrons and $^{15}$N from the chain of the $^{19}$F
production.

To date considerable work has been performed to study the
$^{15}$N($\alpha,\,\gamma$)$^{19}$F reaction
\cite{oli97,wil02,for03,reh12}. As for the
$^{15}$N($n$,\,$\gamma$)$^{16}$N reaction, only two experimental
results \cite{mei96,bar08} were available. In 1996, Meissner et
al. \cite{mei96} measured the $^{15}$N($n$,\,$\gamma$)$^{16}$N
cross section at neutron energies of 25 keV, 152 keV, and 370 keV.
To interpret the results and calculate the
$^{15}$N($n$,\,$\gamma$)$^{16}$N reaction rate, these authors
\cite{mei96} performed direct capture calculations based on the
experimental spectroscopic factors of $^{16}$N from measurement of
the $^{15}$N($d,\,p$)$^{16}$N reaction \cite{boh72} since the
magnitude of the direct capture rate depends directly on the
neutron spectroscopic factors of the four low-lying levels in
$^{16}$N. In that work \cite{boh72} the $^{15}$N($d,\,p$)$^{16}$N
angular distributions were measured by solid-state detector
telescopes with an average energy resolution of 35 keV and yielded
nearly equal spectroscopic factors of $\sim$0.5 for the ground
state and the first three excited states at $E_x$ = 0.120 MeV,
0.298 MeV and 0.397 MeV in $^{16}$N. In 2008, Bardayan et al.
\cite{bar08} measured the angular distributions of the
$^{15}$N($d,\,p$)$^{16}$N reaction in inverse kinematics by using
silicon detector arrays and a recoil separator. Since
closely-spaced levels (ground state + 0.120 MeV level, 0.298 +
0.397 MeV levels) could not be resolved, they used two different
methods to determine the spectroscopic factors of these four
states. The resulting values are larger than the ones in Ref.
\cite{boh72} by a factor of $\sim$2, while are in agreement with
shell model expectations \cite{boh72,mei96} where the four
low-lying $^{16}$N levels are thought to be good single-particle
levels as $^{15}$N is a closed neutron-shell nucleus. Furthermore,
they derived the $^{15}$N($n$,\,$\gamma$)$^{16}$N rate via direct
capture calculations based on their experimental spectroscopic
factors \cite{bar08}. The updated reaction rate in Ref.
\cite{bar08} is nearly twice faster than the ones in Ref.
\cite{mei96} since the spectroscopic factors used in these two
works differ by a factor of $\sim$2.

In short, two previous measurements \cite{boh72,bar08} of the
($d,\,p$) reaction led to the spectroscopic factors with a
discrepancy by a factor of $\sim$2 and further yielded different
$^{15}$N($n$,\,$\gamma$)$^{16}$N rates. Therefore, it is
worthwhile to perform a new measurement of the neutron
spectroscopic factors for the four low-lying $^{16}$N states via
an independent transfer reaction. In this article we report
accurate and precision measurement of the spectroscopic factors
for these $^{16}$N levels via the angular distribution of the
$^{15}$N($^{7}$Li,\,$^{6}$Li)$^{16}$N reaction using a Q3D
magnetic spectrograph. These spectroscopic factors are used to
evaluate the $^{15}$N($n$,\,$\gamma$)$^{16}$N reaction rate.


The experiment was performed at the HI-13 tandem accelerator of
the China Institute of Atomic Energy (CIAE) in Beijing. The
experimental setup and procedures are similar to those previously
reported \cite{guo12,li12,li13}. Melamine
C$_3$N$_3$($^{15}$NH$_2$)$_3$ enriched to 99.35\% in $^{15}$N was
employed as target material with a thickness of 46 $\mu$g/cm$^2$,
evaporated on a 30 $\mu$g/cm$^2$ thick carbon foil. To improve
thermal conductivity of targets metal gold was evaporated on
melamine foil, accumulated to a thickness of 22 $\mu$g/cm$^2$. The
target thickness was determined by using an analytical balance
with a precision of 1 $\mu$g and was verified with the well-known
differential cross sections at $\theta_\mathrm{c.m.}$ =
33.5$^\circ$ and 49.2$^\circ$ of the $^{7}$Li\,+\,$^{15}$N elastic
scattering \cite{woo82,oli96}. An uncertainty of 5\% was assigned
for target thickness, which is reasonable by considering the
balance precision and the error of the well-known differential
cross sections.

A 44 MeV $^{7}$Li beam was delivered to measure the angular
distribution of the $^{15}$N($^{7}$Li,\,$^{6}$Li)$^{16}$N reaction
leading to the ground state and the first three excited states in
$^{16}$N. The angular distribution of the $^{7}$Li\,+\,$^{15}$N
elastic scattering was also measured to obtain the optical model
potential (OMP) for the entrance channel of the transfer reaction.
In addition, a 34.5 MeV $^6$Li beam was delivered for measurement
of the $^6$Li\,+\,$^{15}$N elastic scattering to evaluate the OMP
for the exit channel.

The beam current was measured by a Faraday cup covering an angular
range of $\pm\,$6$^\circ$ in a laboratory frame and used for the
absolute normalization of the cross sections at
$\theta_\mathrm{lab}\,>\,6^\circ$. The Faraday cup was removed
when measuring the cross sections at
$\theta_\mathrm{lab}\,\leq\,6^\circ$. A Si $\Delta E\,-\,E$
telescope located at $\theta_\mathrm{lab}\,=\,25^\circ$ was
employed for the relative normalization of the cross sections at
$\theta_\mathrm{lab}\,\leq\,6^\circ$ by measuring the elastic
scattering of the incident ions on the targets. The reaction
products were analyzed with a Q3D magnetic spectrograph and
recorded by a two-dimensional position-sensitive silicon detector
(PSSD, 50\,$\times$\,50 mm) placed at the focal plane of the
spectrograph. The two-dimensional position information from the
PSSD enabled the products emitted into the acceptable solid angle
to be recorded completely.

\begin{figure}[htbp]
\begin{center}
\resizebox{0.45\textwidth}{!}{
  \includegraphics{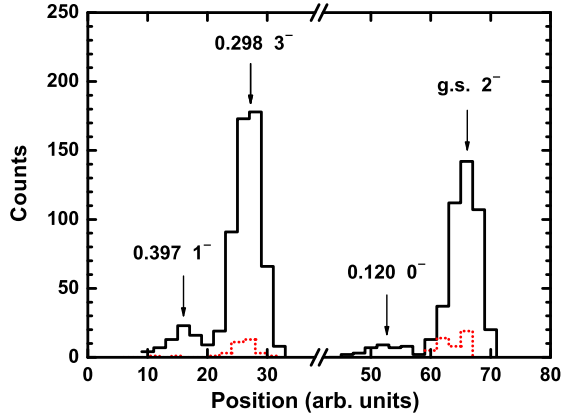}}
\caption{(Color online) Focal-plane position spectrum of the
$^{6}$Li events at $\theta_\mathrm{lab}$ = 10$^\circ$ from the
neutron-transfer reactions. The black solid and red dashed lines
are the results from the enriched $^{15}$N target and natural
$^{14}$N target, respectively. The break in the x-axis denotes the
narrow gap between two separated detectors.} \label{fig1}
\end{center}
\end{figure}

\begin{figure}[htbp]
\begin{center}
\resizebox{0.45\textwidth}{!}{
  \includegraphics{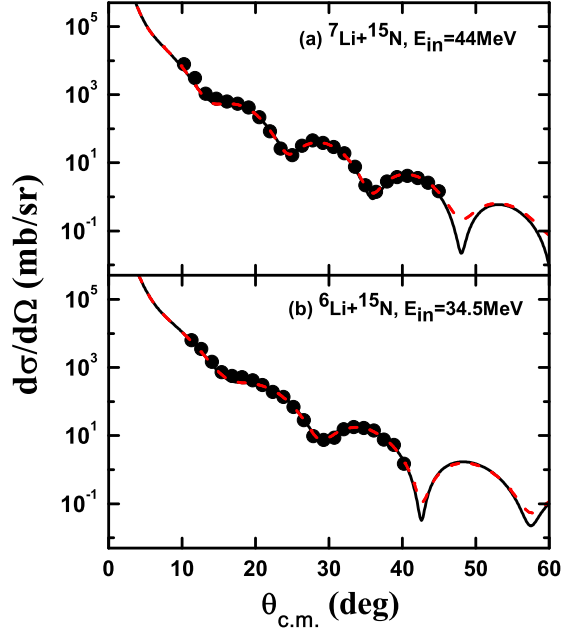}}
\caption{(Color online) Angular distributions of the
$^{7}$Li+$^{15}$N elastic scattering at incident energy of 44 MeV
and the $^{6}$Li+$^{15}$N elastic scattering at incident energy of
34.5 MeV. The black solid and red dashed curves represent the
calculations using the fitted OMP parameters without and with
spin-orbit term, respectively.} \label{fig2}
\end{center}
\end{figure}

\begin{figure*}[htbp]
\begin{center}
\resizebox{0.7\textwidth}{!}{
\includegraphics{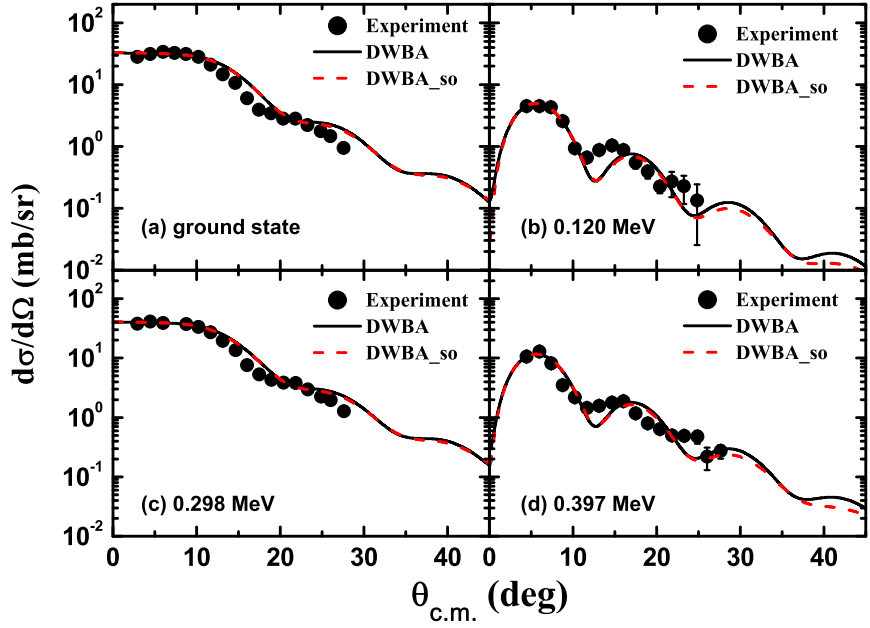}}
\caption{(Color online) Angular distributions of the
$^{15}$N($^{7}$Li,\,$^{6}$Li)$^{16}$N reaction leading to the
ground and first three excited states in $^{16}$N. The black solid
and red dashed curves denote the DWBA calculations with the fitted
OMP parameters without and with spin-orbit term, respectively.}
\label{fig3}
\end{center}
\end{figure*}

As an example, Fig. \ref{fig1} shows the focal-plane position
spectrum of $^6$Li at $\theta_\mathrm{lab}$\,=\,10$^\circ$ from
the neutron-transfer reactions. The background from $^{14}$N is
negligibly small. After background subtraction and beam
normalization, the angular distributions of the elastic scattering
and the $^{15}$N($^{7}$Li,\,$^{6}$Li)$^{16}$N reaction leading to
the ground and first three excited states in $^{16}$N were
obtained, as presented in Figs. \ref{fig2} and \ref{fig3}.

The finite-range distorted wave Born approximation (DWBA) method
with the FRESCO code \cite{tho88} was used to analyze the
experimental angular distributions. The OMP parameters for the
entrance and exit channels were determined by fitting the present
experimental angular distributions of the $^{7}$Li\,+\,$^{15}$N
and $^{6}$Li\,+\,$^{15}$N elastic scattering (Fig. \ref{fig2}).
The starting values of OMP parameters were obtained by fitting the
single-folding nucleus-nucleus potential of Ref. \cite{xu13}. The
real potential was taken to be a squared Woods-Saxon shape, which
fits the real part of the folding model potential better than
usual Woods-Saxon shape does \cite{kho07}. For the imaginary
potential, Woods-Saxon shape was found to be appropriate. In
addition, we investigated the effect of spin-orbit potential
parameters although for heavy ions they are thought to have little
or no influence on the cross sections  \cite{tra00}. Since the
strength of the spin-orbit potential for heavy ions scales as
1/$A$ as compared with the nucleon case \cite{tra00}, the depths
of the entrance and exit channels were derived to be 0.843 MeV and
0.983 MeV using the depth of 5.9 MeV for nucleon \cite{var91}.
Both sets of OMP parameters (with and without spin-orbit
potential) were used to study spectroscopic factors. Full complex
remnant term interactions were included in the transfer reaction
calculations. The parameters of the core-core
($^{6}$Li\,+\,$^{15}$N) potential were obtained using the present
ones of $^{6}$Li\,+\,$^{15}$N at 34.5 MeV and systematics in
energy dependence of the potential parameters \cite{xu13}. For the
bound state $^{16}$N\,=\,$^{15}$N\,+\,$n$, the standard
geometrical parameters $r$ = 1.25 fm and $a$ = 0.65 fm were
adopted, which have been extensively utilized to study the ground
state neutron spectroscopic factors for 80 nuclei of $Z$ = 3-24
\cite{tsa05} and 565 excited state neutron spectroscopic factors
for $Z$ = 8-28 nuclei \cite{tsa09}. All the parameters used in the
calculation are listed in Table \ref{tab1}.

\begin{table*}
\begin{minipage}{15cm}
\begin{center}
\caption{OMP parameters used in the present DWBA calculation.
$E_{\textrm{in}}$ denotes the incident energy in MeV for the
relevant channels, $V$ and $W$ are the depths (in MeV) of the real
and imaginary potentials with squared Woods-Saxon and usual
Woods-Saxon shapes, and $r$ and $a$ are the radius and diffuseness
(in fm). \label{tab1}}
\begin{tabular}{p{2cm}p{1cm}p{1.2cm}p{1cm}p{1cm}p{1cm}p{1cm}p{1cm}p{1cm}p{1cm}p{1cm}p{1cm}}
\hline\hline Channel & $E_{\textrm{in}}$ & $V$ & $r_{v}$ & $a_{v}$
& $W$ & $r_{w}$ & $a_{w}$& $V_{so}$ & $r_{so}$ &
$a_{so}$& $r_{C}$\\
\hline
$^{7}$Li+$^{15}$N & 44.0 & 138.7 & 0.911 & 1.26 & 45.0 & 0.966 & 0.820 & & & & 1.30\\
& & 118.9 & 0.899 & 1.36 & 49.3 & 0.923 & 0.795 & 0.843& 0.974&1.36 & 1.30\\
$^{6}$Li+$^{16}$N & 34.5 & 111.0 & 0.886 & 1.47 & 39.0 & 0.840 & 1.02& & &  & 1.30\\
& & 91.0 & 0.930 & 1.43 & 29.3 & 0.841 & 1.12 & 0.983& 0.786&1.37 & 1.30\\
$^{6}$Li+$^{15}$N & 37.7 & 132.0 & 0.901 & 1.37 & 31.3 & 0.945 & 0.918& & &  & 1.30\\
$n$+$^{15}$N &  & ~~~\footnote{The depth was obtained by fitting
to reproduce the binding energy of the neutron in $^{16}$N.} & 1.25 & 0.65 & & & & 6.0 & 1.25 & 0.65 & 1.25\\
\hline\hline
\end{tabular}
\end{center}
\end{minipage}
\end{table*}

The spectroscopic factors can be derived by the normalization of
DWBA calculation to the experimental angular distribution. The
neutron spectroscopic factor of the $^7$Li ground state needs to
be chosen. So far considerable work has been performed to study
it. Shell model calculations \cite{bar66,coh67,var69,kum74}
predicted values of 0.80, 0.72, 0.79 and 0.77. Three measurements
\cite{li69,tow69,fag76} of the $^7$Li($p,\,d$) reaction yielded
consistent spectroscopic factors of 0.71, 0.72 and 0.87. However,
two measurements \cite{sch67,tsa05} of the $^6$Li($d,\,p$)
reaction derived very different values of 0.9 and 1.85. In
addition, most recent measurement of the elastic-transfer
$^7$Li($^6$Li,\,$^7$Li) reaction yielded a value of 0.73
\cite{su10}. In view of these different evaluations, we decided to
use the value of 0.73 \cite{coh67,li69,tow69,su10}. The 1$p_{3/2}$
and 1$p_{1/2}$ components in the spectroscopic factor of $^7$Li
were taken to be 1.5 based on the shell model calculation
\cite{coh67}. The spectroscopic factors of the ground state and
the first three excited states in $^{16}$N were then extracted to
be 0.96\,$\pm$\,0.09, 0.69\,$\pm$\,0.09, 0.84\,$\pm$\,0.08 and
0.65\,$\pm$\,0.08, respectively. The errors result from the
statistics (8\%, 12\%, 8\%, 11\%), the uncertainty of target
thickness (5\%) and the OMP parameters (1.6\%, 2.2\%, 1.2\%,
3.1\%), respectively. Additional model uncertainties were not
included when evaluating the uncertainty of spectroscopic factors,
as two previous work \cite{boh72,bar08} did. In this work we also
tested the effect of changing the geometrical parameters of the
binding potential on spectroscopic factors. We changed the radius
and diffuseness 20\% around the standard values ($r$ = 1.25 fm,
$a$ = 0.65 fm), and then found that the spectroscopic factors of
the two states corresponding to 1$d_{5/2}$ transfer vary by
$\sim$20\% and those of the two states corresponding to 2$s_{1/2}$
transfer vary by $\sim$8\%. This difference in response to
transfers to the 1$d_{5/2}$ and 2$s_{1/2}$ states may come from
the different peripheralities of these two transitions.

The new spectroscopic factors are listed in Table \ref{tab2},
together with the ones from two previous measurements of the
$^{15}$N($d,\,p$) reaction \cite{boh72,bar08} and the shell model
calculation \cite{mei96}. The present results are approximately
twice larger than those from the $^{15}$N($d,\,p$) reaction
\cite{boh72}, while are in good agreement with those from the
$^2$H($^{15}$N,$\,p$) reaction using Method2 (namely, components
allowed to vary freely) in Ref. \cite{bar08} where two different
methods were used to determine the spectroscopic factors since
closely-spaced levels (ground state + 0.120 MeV level, 0.298 +
0.397 MeV levels) in $^{16}$N could not be resolved. The present
results demonstrate that two levels corresponding to neutron
transfers to the 1$d_{5/2}$ orbit in $^{16}$N are good
single-particle levels but two levels corresponding to neutron
transfers to the 2$s_{1/2}$ orbit are not so good single-particle
levels as the shell model expected \cite{mei96}.

\begin{table*}
\begin{minipage}{15cm}
\begin{center}
\caption{The present spectroscopic factors of $^{16}$N and other
available results in the literature. $nl_j$ is single-particle
shell quantum number. \label{tab2}}
\begin{tabular}{p{1.5cm}p{1cm}p{1cm}ccp{2cm}p{2cm}c}
\hline \hline $E_{x}$& $J^\pi$ & $nl_j$ & \multicolumn{5}{c}{Spectroscopic factor}\\
\cline{4-8}
(MeV)&&&OXBASH~\cite{mei96}&~$^{15}$N($d,\,p$)~\cite{boh72}\footnote{30\%
uncertainty for their spectroscopic factors.}~~&\multicolumn{2}{c}{$^2$H($^{15}$N,$\,p$)~\cite{bar08}~~~~~}&Present\\
&&&&&Method1\footnote{The individual components were weighted by
the relative spectroscopic factors reported in Ref.
\cite{boh72}.}& Method2\footnote{The magnitudes of the individual
components were
allowed to vary freely in fit.}&\\
\hline
0 & 2$^-$ & 1$d_{5/2}$ & 0.93 & 0.55 & 0.96(15) & 1.04(16) & 0.96(9)\\
0.120 & 0$^-$ & 2$s_{1/2}$ & 0.95 & 0.46 & 0.80(12) & 0.71(12) & 0.69(9)\\
0.298 & 3$^-$ & 1$d_{5/2}$ & 0.87 & 0.54 & 0.91(14) & 1.03(16) & 0.84(8)\\
0.397 & 1$^-$ & 2$s_{1/2}$ & 0.96 & 0.52 & 0.88(13)& 0.74(12)&0.65(8)\\
\hline \hline
\end{tabular}
\end{center}
\end{minipage}
\end{table*}

\begin{figure}[htbp]
\begin{center}
\resizebox{0.45\textwidth}{!}{
  \includegraphics{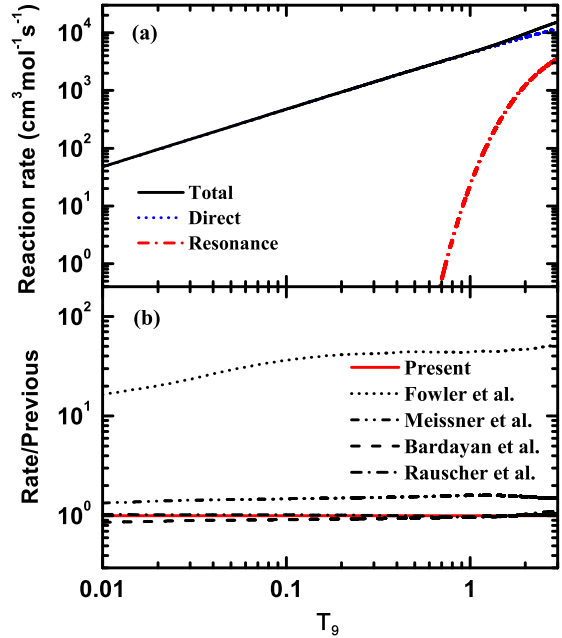}}
\caption{(Color online) The present $^{15}$N($n,\,\gamma$)$^{16}$N
rate for temperatures of 0.01-3 GK, and comparison of the present
rate with the previous results by Fowler et al. \cite{fow67},
Rauscher et al. \cite{rau94}, Meissner et al. \cite{mei96} and
Bardayan et al. \cite{bar08}. $T_9$ is the temperature in units of
1 GK. See text for details.} \label{fig4}
\end{center}
\end{figure}

We computed the direct capture cross section and the rate of
$^{15}$N($n,\,\gamma$)$^{16}$N based on the measured spectroscopic
factors using the RADCAP code \cite{ber03}. At low energies of
astrophysical interest, the $^{15}$N($n,\,\gamma$)$^{16}$N direct
capture cross section is dominated by the $E$1 transition from
incoming $p$-wave to the bound state. Note that the same
geometrical parameters need to be used when calculating the bound
state wave function as those used for deriving the spectroscopic
factors of $^{16}$N. In addition, the parameters for computing the
scattering wave function are identical to the ones for the bound
state potential, as suggested by Huang et al. \cite{hua10}. We
also evaluated the contribution of the resonant capture via the
state at $E_x$ = 3.523 MeV using the resonance parameters given by
Meissner et al. \cite{mei96}. The resulting rates are shown in
Fig. \ref{fig4}(a). The rate is dominated by the direct capture at
temperatures of $T_9$\,$<$\,2. The total reaction rate at $T_9$ =
0.1 (typical temperature of AGB stars) was found to be
504\,$\pm$\,66 cm$^3$mol$^{-1}$s$^{-1}$, The error results from
the uncertainty (12\%) of the present spectroscopic factors and
that (5.3\%) of the geometrical parameters.

In Fig. \ref{fig4}(b) we compare the present rate with the
previous ones available in the literature
\cite{fow67,rau94,mei96,bar08}. The present rate is about 20-50
times faster than the previous calculation \cite{fow67} since it
is a first order estimate of the $p$-wave capture contribution.
The present rate is also nearly twice faster than the experimental
ones in Ref. \cite{mei96}. This is because the present
spectroscopic factors are larger than the ones from the
$^{15}$N($d,\,p$) reaction \cite{boh72} used in Ref. \cite{mei96}.
In addition, the new rate within 10\% agrees with the ones in
Refs. \cite{rau94,bar08} since the new spectroscopic factors are
consistent with those from the shell model calculation
\cite{rau94} and the $^2$H($^{15}$N,$\,p$) reaction \cite{bar08}.

The new rate was fitted with the expression used in the
astrophysical reaction rate library REACLIB \cite{thi87},
\begin{eqnarray}
\label{eq1}%
N_A \langle\sigma v\rangle =
\exp[6.38075+0.0159372T_{9}^{-1}\nonumber\\-1.80122T_{9}^{-1/3}
+4.02273T_{9}^{1/3}-0.108521T_{9}\nonumber\\-0.0221989T_{9}^{5/3}
-0.754253\ln(T_{9})].
\end{eqnarray}
The overall fitting errors are less than 6\% at temperatures from
0.01 to 10 GK.


In summary, the angular distributions of the
$^{15}$N($^{7}$Li,\,$^{6}$Li)$^{16}$N reaction leading to the four
low-lying $^{16}$N levels were measured and used to derive their
spectroscopic factors. The closely-spaced levels in $^{16}$N were
resolved and the accuracy and precision of spectroscopic factors
are enhanced due to the first application of high-precision
magnetic spectrograph for study of the neutron transfer reaction
on $^{15}$N. The present work demonstrates that two levels
corresponding to neutron transfers to the 1$d_{5/2}$ orbit in
$^{16}$N are good single-particle levels but two levels
corresponding to neutron transfers to the 2$s_{1/2}$ orbit are not
so good single-particle levels as the shell model expected
\cite{mei96}.

We also derived the $^{15}$N($n,\,\gamma$)$^{16}$N reaction rate
using the measured spectroscopic factors. The present rate is
about 20-50 times faster than the previous calculation
\cite{fow67} and is nearly twice faster than the experimental ones
in Ref. \cite{mei96}. In addition, the new rate is within 10\% in
agreement with the ones in Refs. \cite{rau94,bar08}. The present
work provides an independent examination to shed some light on the
existing discrepancies in the spectroscopic factors of the four
low-lying $^{16}$N states and the stellar rate of the
$^{15}$N($n$,\,$\gamma$)$^{16}$N reaction.

The authors thank the staff of HI-13 tandem accelerator for the
smooth operation of the machine, R. C. Johnson and N. K. Timofeyuk
for their helpful discussions. We acknowledge the anonymous
referee for helpful comments and suggestions. This work was
supported by the National Natural Science Foundation of China
under Grant Nos. 11075219, 11321064, 11375269, 11275272 and
11275018, the 973 program of China under Grant No. 2013CB834406.

\end{CJK*}

\end{document}